# 16
# AUTOMATA DESCRIBING OBJECT BEHAVIOR

Bernhard Rumpe and Cornel Klein
*Institut für Informatik, Technische Universität München*
*80333 München, Germany*
*http://www4.informatik.tu-muenchen.de/*

## ABSTRACT

Relating formal refinement techniques with commercial object-oriented software development methods is important to achieve enhancement of the power and flexibility of these software development methods- and tools. We will present an automata model together with a denotational and an operational semantics to describe the behavior of objects. Based on the given semantics, we define a set of powerful refinement rules, and discuss their applicability in software engineering practice, especially with the use of inheritance.

## 1 INTRODUCTION

### 1.1 Software Development Methods: Theory and Practice

In the industrial practice of software engineering in the last fiveteen years a multitude of so-called software development methods have been developed. Such methods, such as SSADM [AG90], OMT [RBPEL91] or Fusion [CABDGHJ94], normally use different description techniques for describing different views of a software product to be developed. On the one hand, these description techniques provide notations which are well-suited for the communication with the application expert, and which can be efficiently used for typical modeling problems such as data modeling (e.g. entity-/relationship-diagrams, [C76]). On the other hand, however, these description techniques lack a precisely defined semantics. Even their syntax is sometimes only defined informally, e.g. by giving examples. As a result, a lot of problems during the application of such methods arise, which are caused by the ambiguous interpretation of the semantics of the used modeling concepts:

- no description techniques are provided for typical modeling problems such as e.g. data modeling.

- the refinement calculus provides only a set of "low-level", fine-grained refinement rules with an emphasize on completeness. What is needed are powerful, large-grained refinement rules which are tailored for typical refinement problems and typical description techniques.

It is the aim of this paper to present a step towards closing the above-sketched gap between theory and practice. We believe that both worlds, theory and practice, can benefit from such an approach. For formal methods, commercially successful description techniques may aid in scaling up these methods for their application in the development of larger systems. Commercial methods may benefit conceptually by more precisely defined concepts, and by CASE-tools with enhanced functionalities such as improved consistency checks and more powerful generation- and analysation facilities.

## 1.2  Overview of the paper

Typical for all object-oriented software development methods are description techniques of the following two kinds:

- A description technique for *data modeling*.
  These description techniques are normally based on the entity-/relationship-model [C76], which is extended in various ways. Data models describe the structure of the persistent data of a system, and some integrity constraints. Entity-/relationship-modeling is the most widely used modeling technique.

- A description technique for modeling *behavior*.
  State-Transition diagrams, or automata, are used to model the behavior of the whole system, of subsystems or of single objects. For reducing the complexity of the notation, hierarchical state-transition diagrams such as statecharts [H87] have been proposed, which are used e.g. in OMT.

In this paper, we will present an automaton model which is well-suited for the description of the behavior of *objects*. We will consider objects as being active entities encapsulating a local data space and a local process, and which communicate with *asynchronous message passing*. This kind of objects can not be tackled adequately with well known modeling approaches, but they are gaining importance in industry in the context of distributed client-/server applications.

The paper is organized as follows: In the following section, we define the abstract syntax and a concrete graphical notation for our automata. The main purpose of the graphical notation is to illustrate the examples. In section 3, the denotational semantics of an automaton is given by a predicate characterizing a set of stream processing functions (see [BDDFGW93], [RKB95]), and the operational semantics is defined as the set of their executions. Based on the given denotational semantics, we will continue by define a calculus for refinement which is well-suited for the use by application experts and software engineers. We show the usefulness of the refinement rules by a short example.

## 2  MESSAGE PROCESSING AUTOMATA

In this section, we define the abstract syntax for message processing automata (in short automata) as well as one notation for them.

We will use automata to model the input-/output behavior of components. For this purpose, we will not use a black-box-view, which only relates the input and the output of a component. Instead, we will use an abstraction of the internal structure of a component. This approach is also called the "state-box-view", e.g. in the cleanroom software engineering method ([MDL87], [HM93]). We will also abstract from the concrete messages which objects may receive and send. Instead, we model equivalence classes of these messages, called *characters*. One such equivalence class is for example the set of all messages of the same type, abbreviated by the name of this type.

An automaton describes the reaction of a component with respect to a given input stimulus and a given state. The automaton consumes the input message, produces a sequence of output messages, and enters a new state for processing the next input message. Thus, transitions are labeled with sequences of messages, as this is the case in mealy automata [HU90].

Our approach for modeling the input-/output behavior of components is closely related with I/O-automata ([J87], [J85], [LS89]). One transition of an I/O-automaton can input a character, output a character, or it can be an internal transition. The processing of an input character triggers a sequence of further transitions, where each transition may output a further character. To guarantee that the environment of an I/O-automaton may send messages at any point in time, an additional constraint called "input enabledness" is imposed on the transition relation. Because an I/O-automaton can only accept or emit one character per transition, it needs a fine grained set of control states. Our notion of an automaton is more abstract, it labels transitions with the input character as well as with output characters, thus modeling the input characters causal for the output. Therefore, no intermediate control states are necessary. One transition models the handling of a complete message, describing input and output as well as the internal state change. We only use abstract data states, characterizing equivalence classes of the data states of the modeled components.

## 2.1 Abstract Syntax for Automata

The abstract syntax contains only the essence of the concrete textual or graphical notation of the automaton, while it ignores the keywords used or the concrete shape of the graphical symbols. We define the abstract syntax of an automaton as follows:

**Definition 1 (Message Processing Automaton)**
*A message processing automaton is a 4-tuple (S,M,$\delta$,I), consisting of*

- *a set of states S,*

- *a set of input and output characters M,*

- *a state transition relation $\delta \subseteq$ S×M×S×M$^*$, where the first two components are the source state and the input message and the second two components are the destination state and a sequence of output messages, and*

- *a set I $\subseteq$ S×M$^*$ of pairs each consisting of a start state and an initial output.*

**(End Definition)**

I will be called the set of initial elements. Instead of (s,m,t,out)$\in\delta$ we will often write $\delta$(s,m,t,out). For given source s and input m, we write $\delta$(s,m) as a shorthand if there exists a destination t and output out such that $\delta$(s,m,t,out).

An automaton models the input-/output behavior and the state changes of an object, which processes messages from the set M. A sequence of characters is processed in the following way: Nondeterministically, according to the first character of the input stream and the current state of the automaton a matching transition from $\delta$ is selected. The transition is labeled with output characters, which are sent out during the processing of the input character. The automaton enters a new state according to the transition, in which he continues the processing of the input stream where the first character has been removed.

A transition models the fact that the output of the transition causally depends on the input of the transition. During the processing of a transition, the output is emitted. This does not mean that the output immediately follows the input, but only that the output is caused by the input, and is sent sometimes later. This corresponds to possible message delay between distributed, asynchronously communicating objects.

In contrast to finite automata ([T90], we also allow infinite state sets, and extend transitions with output. We do not use final states, because we do not model terminating components, but components with an infinite lifetime.

**Definition 2 (Total Automaton)**
*An automaton (S,M,$\delta$,I) is called* total, *if for each input in each state at least one transition exists:* $\forall$s$\in$S,m$\in$M. $\delta$(s,m)$\neq\emptyset$. *An automaton which is not total is called* partial.

**(End Definition)**

## 2.2   Notation for Automata

To illustrate our automata model, in the sequel we will use a graphical representation for automata. Nodes will represent (equivalence classes of) states of an automaton, while directed arcs will be used to represent the transitions of an automaton. Another possibility for the concrete syntax of automata might be the use of tables, as this is the case in [J93a], [P92] and [S94]. We will not discuss the various advantages of graphical and textual notations for software engineering here. An excellent survey concerning this topic can be found in [P95].

**Example 1 (Parity automaton)**
  *The automaton in figure 1 describes the behavior of an object computing the parity of its input messages. The automaton has two states, representing an even or odd sum of the bits received so far. The current parity can be requested by issuing a* ?.

**(End Example)**

| Parity: | $M = \{0, L, ?\}$ | |
|---|---|---|
| 0/<br>?/0 | L/ | 0/<br>?/L |
| /<br>stdPriority.ps | L/ | |

Figure 1: Graphical representation of the parity automaton

   In general, the nodes represent the states S. The state transition relation $\delta$ is given by arcs, which are labeled with one input character m and with an output sequence out in the form m/out. The empty output sequence $\varepsilon$ is omitted for simplicity. The initial states are characterized by arcs without source node, which are labeled with the initial output in the form /out. Only the set $M$ of input- and output characters has to be given explicitly.
   In software engineering methods like [SM92], [B94], [CABDGHJ94] or [J93] state transition graphs or hierarchical extension of these are used to represent the behavior of components. Due to the fact that the state space of components is infinite in general, in these methods one node of a graph represents an equivalence class of states. To allow for the formulation of propositions about states, pre- and postconditions are used. We will illustrate this by specifying an object which realizes an unbounded FIFO-buffer (see figure 2). Transitions are additionally labeled by pre- and postconditions in a suitable formal language. For the pre- and postconditions we use the well-known Hoare notation. The variable `s` denotes the source state and the variable `s'` denotes the target state of a transition. The functions $\#$, *ft* and *rt* are defined in the appendix. Because one node of the graph represents a non-empty set of states, there has to be a mapping from nodes in the graph to sets of states. This mapping is defined by conditions which are attached to the nodes and which constitute a partition of the state space.

```
    Buffer:       M = D ∪ {?},   S = D*,  I = {(ε,ε)}
```

|  |  |  |
|---|---|---|
| ?/? | d/ {s'=<d>} | d/ {s'=s^d}<br>{#s≥2} ?/ft(s) {s'=rt(s)} |
| s=ε |  | s≠ε |
| /<br>stdPuffer.ps | {s=<d>} ?/d |  |

Figure 2: Graphical representation of a buffer automaton

# 3  SEMANTICS FOR AUTOMATA

We now give a denotational semantics for our automata. The denotational semantics associates a set of stream processing functions with each automaton. Stream processing functions provide an abstract, compositional semantics for asynchronous communication objects [BDDFGW93]. We will also give an operational semantics, associating a set of transition sequences with each automaton. These transition sequences reflect the intuitive understanding of the operational behavior of an automaton.

## 3.1  Denotational Semantics

The semantic model of stream processing functions is introduced in the appendix. The semantics of an automaton is a set of stream processing functions.

A set of stream processing functions can either be viewed to model an *underspecified* agent, i.e. as an agent in the specification of which some details have been left open, or as a model of a *non-deterministic* agent, i.e. as an agent which non-deterministically chooses between alternatives during its lifetime. The difference between nondeterminism and underspecification can not be observed.

Non-determinism in our automata occurs because of the non-deterministic choice of the transition relation $\delta$. Here, any transition with matching source state and input character may be chosen:

**Definition 3 (Semantics for Total Automata)**
*The semantics of a total automaton* $(S,M,\delta,I)$ *is defined as follows:*

$$[\![(S,M,\delta,I)]\!]^c \stackrel{def}{=} \{\ g \in M^\omega \stackrel{s}{\to} M^\omega \mid$$
$$\exists\ h \in [\![(S,M,\delta,I)]\!]^C,\ (s_i, out_i) \in I.\ \forall in.\ g(in) = out_i\hat{\ }h(s_i, in)\ \}$$

*where* $[\![.]\!]^C$ *is the greatest set of state parameterized functions satisfying the following equation (the greatest fixpoint via set inclusion):*

$$[\![(S,M,\delta,I)]\!]^C \stackrel{def}{=} \{\ h \in S \times M^\omega \stackrel{s}{\to} M^\omega \mid (\forall s.\ h(s,\varepsilon) = \varepsilon)\ \wedge$$
$$\forall m, s. \exists t, out. \delta(s,m,t,out) \wedge \exists h' \in [\![(S,M,\delta,I)]\!]^C.$$
$$\forall in.\ h(s, m\hat{\ }in) = out\hat{\ }h'(t, in)\}$$

**(End Definition)**

The recursively defined set $[\![(S,M,\delta,I)]\!]^C$ consists of state parameterized stream processing functions. According to a given input message `m` and the current state `s` non-deterministically a transition `(s,m,t,out)` is chosen, the output `out` is emitted and the new state `t` is entered. To allow for maximum non-determinism, a new function `h'` is chosen to model the behavior in the new state. For empty input streams, an empty output stream is emitted.

In [R96] it has been shown that the

- semantics of total automata is well-defined,

- that automata cannot be inconsistent in the sense that they denote an empty set of stream processing functions, if set `I` is nonempty.

A partial automaton can be viewed as shorthand for an automaton which in certain cases leaves the target state and the output completely unspecified. Such an automaton allows for any behavior in cases where the state transition relation is partial for certain inputs. We call this situation *chaos*. Partial automata can easily be totalized by adding auxiliary transitions. This way, the semantic definition for partial automata can be reduced to the semantic definition for total automata.

### 3.2 Operational Semantics

We now define the operational semantics for our automata, as a set of executions. An execution describes the transitions of an automaton which are used during the processing of a certain input sequence. An execution also describes which output is produced and how the output causally depends on the input.

An execution is composed of the following components, which describe so-called *execution elements*:

$s \xrightarrow{in/out} t$ is a (normal) transition with output, when $\delta(s,in,t,out)$.

$\xrightarrow{out_0} s_0$ describes an initial element, when $(s_0, out_0) \in I$.

Formally the execution elements are members of the sets $(S \times M \times S \times M^*)$ and $(S \times M^*)$.

**Definition 4 (Executions)**
*An* execution *is a finite or infinite sequence of execution elements. The first element of an execution has to be an execution element without source state, while all other elements have to be execution elements corresponding to transitions. The source- and target states of subsequent execution elements have to be identical. It is depicted as:*

$$\xrightarrow{out_0} s_0 \xrightarrow{in_1/out_1} s_1 \xrightarrow{in_2/out_2} s_2 \cdots$$

**(End Definition)**

Each execution describes the set of the traversed states as well as the processed input and the produced output during execution of the transitions. The operational semantics of an automaton is the set of all possible executions.

In [R96] it has been shown that the denotational and operational semantics of our automata correspond to each other. The theorem proven shows that the intuitive understanding of the computations, which is formally modeled using the operational semantics, corresponds to the denotational semantics, which is better suited for the correctness proof of the refinement rules in the following section.

## 4 REFINEMENT TECHNIQUES

Refinement techniques are a necessary prerequisite for efficient software production, for reusing given components, and for a transformational software development starting with very abstract, underspecified components and resulting in concrete and efficient executable code. Another area of application is the inheritance of behavior from a super-class to a sub-class. This has been studied extensively in another context in [PR94].

## 4.1  The Refinement Calculus

For the tractability of a refinement calculus it is important that the transformation rules are described on the syntactic level. Nevertheless, the transformation rules have to have a well defined underlying mathematical semantics for ensuring their correctness. The refinement relation we use at the semantic level is the inclusion relation between sets of stream processing functions. This way, refinement corresponds to the reduction of underspecification (or non-determinism).

We define refinement as follows:

**Definition 5 (Refinement)**
*An automaton* (S',M,$\delta$',I') *is called* refinement *of the automaton* (S,M,$\delta$,I), *iff*

$$[\![(\mathtt{S}',\mathtt{M},\delta',\mathtt{I}')]\!] \subseteq [\![(\mathtt{S},\mathtt{M},\delta,\mathtt{I})]\!].$$

*This refinement relation is denoted by*

$$(\mathtt{S},\mathtt{M},\delta,\mathtt{I}) \leadsto (\mathtt{S}',\mathtt{M},\delta',\mathtt{I}')$$

**(End Definition)**

Note that the refinement relation is a transitive relation due to the fact that the inclusion between sets (of stream processing functions) is transitive.

Transformation rules often need additional constraints (so called *application conditions*) which have to be fulfilled to ensure that a transformation rule can be successfully applied. While from the theoretical point of view a complete and powerful set of refinement rules might be desirable, from the practical point of view it is more important that these conditions can be effectively checked.

The transformation rules given in figure 3 are very elementary. Their full power only reveals by their adequate composition to more powerful transformation rules.

These rules can be understood as follows:

**(Arb)**  allows to start a development process by creating a new automaton with arbitrary state set, message set, transition relation and initial states.

**(RemI)**  allows to refine an existing automaton by removing initial states thus reducing the initial choice and thus nondeterminism.

**(RemT)**  allows for the removal of transitions if alternative transitions exist, also reducing nondeterminism.

**(AddT)**  allows the addition of transitions if so far no corresponding transitions exist. Therefore the automaton gets more robust, because chaotic behavior is replaced by an explicit description of behavior.

| | | |
|---|---|---|
| (Arb) | $$\frac{\quad t \quad}{(\mathtt{S},\mathtt{M},\delta,\mathtt{I})}$$ | |
| (RemI) | $$\frac{(\mathtt{S},\mathtt{M},\delta,\mathtt{I})\ ^t}{(\mathtt{S},\mathtt{M},\delta,\mathtt{I'})}$$ | $\mathtt{I'} \subseteq \mathtt{I}$ |
| (RemT) | $$\frac{(\mathtt{S},\mathtt{M},\delta,\mathtt{I})\ ^t}{(\mathtt{S},\mathtt{M},\delta',\mathtt{I})}$$ | $\delta' \subseteq \delta$<br>$\forall \mathtt{s}\in\mathtt{S},\mathtt{m}\in\mathtt{M}.\ \delta(\mathtt{s},\mathtt{m}) \Rightarrow \delta'(\mathtt{s},\mathtt{m})$ |
| (AddT) | $$\frac{(\mathtt{S},\mathtt{M},\delta,\mathtt{I})\ ^t}{(\mathtt{S},\mathtt{M},\delta',\mathtt{I})}$$ | $\delta \subseteq \delta'$<br>$\forall \mathtt{s}\in\mathtt{S},\mathtt{m}\in\mathtt{M}.\ (\delta'\backslash\delta)(\mathtt{s},\mathtt{m}) \Rightarrow \neg\delta(\mathtt{s},\mathtt{m})$ |
| (RemS) | $$\frac{(\mathtt{S},\mathtt{M},\delta,\mathtt{I})\ ^t}{(\mathtt{S}',\mathtt{M},\delta',\mathtt{I})}$$ | $\mathtt{reach}(\mathtt{S},\mathtt{M},\delta,\mathtt{I}) \subseteq \mathtt{S}' \subseteq \mathtt{S}$<br>$\delta' = \delta \cap \mathtt{S}'\times\mathtt{M}\times\mathtt{S}'\times\mathtt{M}^*$ |
| (AddS) | $$\frac{(\mathtt{S},\mathtt{M},\delta,\mathtt{I})\ ^t}{(\mathtt{S}',\mathtt{M},\delta,\mathtt{I})}$$ | $\mathtt{S} \subseteq \mathtt{S}'$ |
| (RefS) | $$\frac{(\mathtt{S},\mathtt{M},\delta,\mathtt{I})\ ^u}{(\mathtt{S}',\mathtt{M},\delta',\mathtt{I'})}\ v$$ | $\alpha\colon \mathtt{S}' \to \mathtt{S}$ total, surjective<br>$\delta'=\{(\mathtt{s},\mathtt{m},\mathtt{t},\mathtt{out}) \mid \delta(\alpha\ \mathtt{s},\mathtt{m},\alpha\ \mathtt{t},\mathtt{out})\}$<br>$\mathtt{I'} = \{(\mathtt{s}',\mathtt{out}) \mid (\alpha(\mathtt{s}'),\mathtt{out})\in\mathtt{I}\}$ |

Figure 3: Refinement rules

**(RemS)** allows for the removal of non-reachable states, where `reach` denotes the set of reachable sets of an automaton.

**(AddS)** allows for the addition of new states.

**(RefS)** allows for the refinement of states. This way a single state can be

　　　　refined into a more fine grained set of states.

As already mentioned, for the practical applicability of a refinement calculus it is important that the applicability conditions of the refinement rules can be checked automatically. If the state set and the transition set are both final, as this is the case in description techniques for software engineering methods, all the applicability conditions of all refinement rules can in fact be checked automatically. However, the situation gets more complex if one uses a finite representation of an automaton which has infinite state- and transition sets, as this was the case in section 2.2. We will not study this further, however one soon gets problems with the decidability of the applicability conditions. See also [PR94] and [PR94b].

The refinement rules given in figure 3 leave the syntactic interface of a refined component unchanged. In [R96] an extension to refine the interface of an object is also given. It allows for extending the input and output set of characters as well as refining one abstract character (such as a message name) by a set of characters (such as the set of possible messages). It also may be used to add further components to messages arriving at or emitted from an object. Especially object-identifiers for identifying the receiver of a message may be added. Thus object-identifiers may be left out if the abstract behavior of an object should be modeled and only later be introduced if the object are modeled more concrete for implementation purposes.

### 4.2  Refinement Example

In order to demonstrate the usefulness of our calculus, we now show a small example for a development process where the behavior of objects is refined step by step by the presented refinement calculus. In figure 4 the development process of the behavior of objects of a class Figure is shown. In figure 5 the continued development for subclass 2D−Figure is depicted. The development is somewhat erratic (as it often is in practice), because it uses intermediate development steps that do not contribute to the result, but are undone by other development steps. This is due to our intension of showing all kinds of refinement steps, their flexibility and their combined application within one example. For simplicity, we do not model output within this example. However, it would be no problem to add output restrictions at any stage of the development process.

Objects of class Figure represent objects shown on the screen. These objects may be selected and deselected by sending appropriate messages to them. The development process shown in figure 4 of Figure-objects consists of the following steps, corresponding to applications of the rules of our calculus:

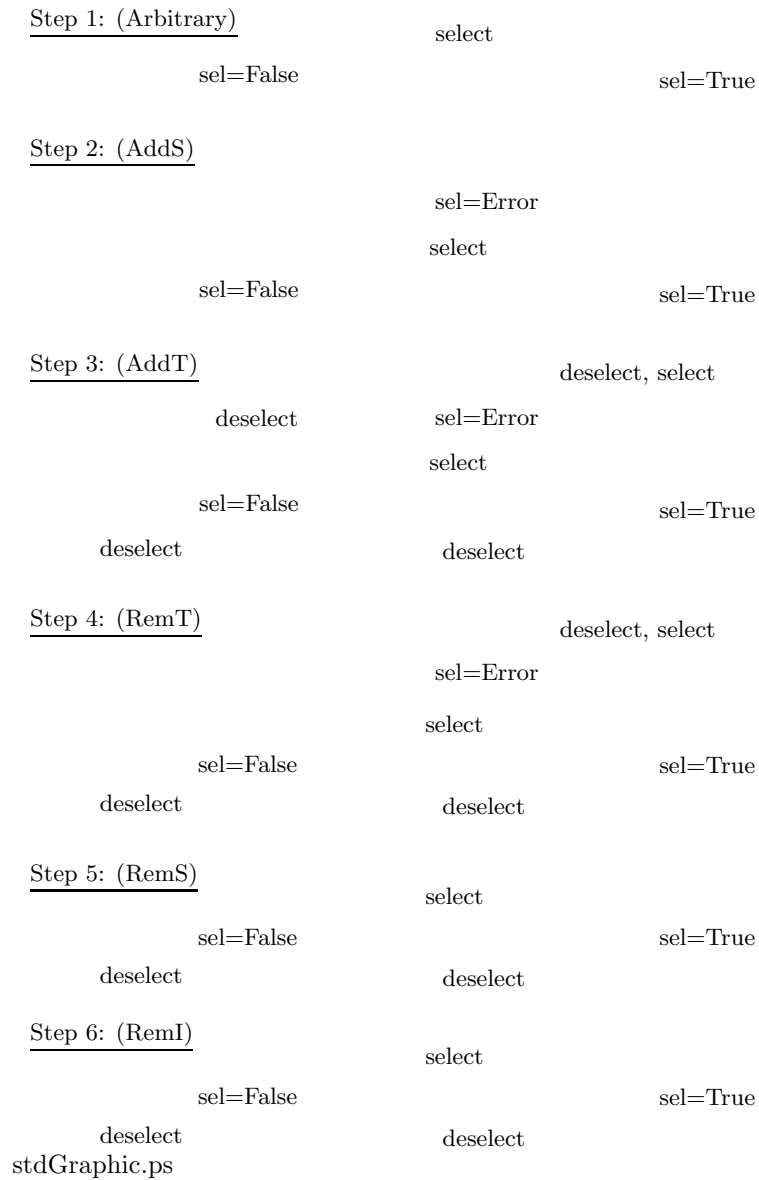

Figure 4: Development of behavior of class `Figure`

**Step 1** : At the beginning of our development, we start with a very simple automaton: It has two states reflecting a selected and an unselected figure. It has one transition modeling what happens if the figure object is sent a `select` if it is not yet been selected. Both states are initially allowed.

**Step 2** : We want to add transitions for a deselection, but what happens if the figure is already deselected? We decide to introduce an additional state, reflecting this error.

**Step 3** : We now add `deselect`-transitions, leaving open whether `deselect` results in an error or is just being ignored in the deselected state.

**Step 4** : The customer wants a robust implementation of `deselect`. Therefore, we remove the possibility to result in an error. Another alternative would be to require the output of a warning message in the the left *deselect*-loop.

**Step 5** : The previously introduced error state is now superfluous and can be removed.

**Step 6** : In a last development step we decide that a newly introduced figure is automatically selected and therefore remove one initial element.

The development for class `Figure` is now finished. It describes the behavior of any object of this class, in terms of the state change according to a given sequence of input messages.

The substitutability principle ([WZ88], [W90]) for objects now enforces the inheritance of this automaton to all subclasses of `Figure`. This not only means that the signature is preserved or extended, but that behavior is inherited in some way. An automaton thus can be seen as an interface description which may be viewed as a contract on which clients may rely on.

We now continue the development process by developing a class `2D−Figure`, which is a subclass of `Figure`. `2D−Figure` in addition allows to fill and empty its area. We start with the inherited automaton from class `Figure`:

**Step 7** : We add transitions for filling end emptying selected 2D-figures. The figures remain selected.

**Step 8** : Filling changes the state of a 2D-figure. This is not modeled. Therefore, we refine the in which a figure is selected into two states. This has the effect, that every transition with source or destination in the unrefined state is duplicated. This way, underspecification has been introduced.

**Step 9** : We now remove some transitions to describe the behavior of `fill` and `empty` in more detail. In addition we model that a newly created 2D-figure is not filled.

Step 7: (AddT)

                                select

    sel=False                                   sel=True

deselect             deselect                 fill, empty

Step 8: (RefS)                                 select

                        deselect

                                         sel=True
    sel=False                                   no contents
                        deselect
deselect                           fill, empty     fill, empty

       select                             fill, empty
                        sel=True
                        filled cont.

                             fill, empty

Step 9: (RemT) and (RemI)              select

                        deselect
                                        sel=True
    sel=False                                   no contents
                        deselect
                              fill
deselect                                             empty

       select                             empty
                        sel=True
                        filled cont.

                             fill

stdGraphic2.ps

Figure 5: Development of behavior of class 2D—Figure

The presented development steps should be sufficient to show that the refinement calculus can be applied to real application development problems, and that the rules of our calculus exactly reflect the development steps used for the development of behavior descriptions. However, this small example also shows that a rigid tool support is inevitable for larger example.

## 5   RELATED WORK

Recently, various approaches for formalizing methods of systems and software development were given. Well known are the so-called "meta-models", originating in the context of tool integration, (see [T89] and [HL93]). However, by this "models" almost only the abstract syntax of the description techniques is captured. An overview of several projects concerning the integration of struc-tured methods with techniques of formal specification can be found in [SFD92]. In [H94], the British standard method SSADM [AG90] is formalized using the algebraic specification language SPECTRUM [BFGHHNRSS93]. This work is continued in the project SYSLAB [B94a]. It is the aim of this project to provide a scientifically founded approach for software- and system development. SYSLAB emphasizes the early phases (analysis, requirements definition, logical design), prototyping, reuse, and code-generation. In the context of the SYSLAB-project, a multitude of description techniques have been formalized up to now. Examples are entity-/relationship-diagrams ([H93], [H95]), data flow techniques [N93a] or time sequence diagrams [F95]. The formalization in most of these cases is based on a so-called mathematical system model or on FOCUS, which have been presented in [RKB95] and [BDDFGW93]. This mathematical system model, which is based on streams and stream processing functions, is also the basis for this paper, and it is presented in the appendix.

The specialization of automata has been a research topic of some groups, especially in the context of object oriented systems.

In software engineering methods ([RBPEL91], [CABDGHJ94] oder [SM92]) automata are used for the behavior modeling of systems and its components. Due to the informal semantics of the description techniques used in these methods in general no guidance is given for determining the relationship of automata of classes being in an inheritance-relationship.

Closely related to our approach is the work presented in [PR94]. However, in [PR94] output actions are not considered and only one (powerful) refinement rule is given. The semantic model of stream processing functions is not used. First order formulae are used for expressing the pre- and postconditions of transitions.

The work in [LW93], [AC93], [C93] is based on the basic principle of *substitutability*, which has been presented in [WZ88] and [W90]. Here, substitutability means that elements (objects) of a supertype can safely be substituted by objects of a subtype. These approaches also study the specialization of behavior in the context of subtyping, which is interesting if the behavior of methods of objects is specialized by inheritance. The concepts developed in these approaches are partially integrated in some object-oriented languages, e.g. in Modula-3 [CDGJKN92], [A93].

Nierstrasz uses finite automata in [N93] for typing objects. The automata are used to model which messages are accepted by an object in which states. This way, the type of an object not only contains information about the static method interface, but also information about the sequences of possible method invocations of a client. Moreover, Nierstrasz defines a subtype relation which corresponds to a behavior specialisation. The subtype-relation is suitable for sequential as well as for distributed, parallel communication. However, the construction of the subtype-relations gets very complex, possibly due to the underlying synchronous communication paradigm which is based on Milner's CCS [M89].

A simpler refinement calculus which is based on asynchronous communication and an I/O-automaton approach is given in [BHS96]. There, input- and output characters are not distinguished and only one character is allowed for each transition. The semantics and the refinement rules are based on a trace semantics.

In [PR94b] we studied how automata can be integrated in the algebraic specification language SPECTRUM [BFGHHNRSS93]. There, an automaton is viewed as a special notation for a logical axiom. This way, a visual description technique is integrated in an axiomatic specification language, while the power and flexibility of SPECTRUM is available in cases where automata can not be used adequately. Examples where SPECTRUM can be applied are pre- and postconditions, as they have also been used in this paper.

The translation process of automata, which have been developed in the design phase, in an implementation using current object-oriented programming languages like C++ [S91] is error-prone. Therefore, in [R94] we used finite automata the transitions of which are labeled with executable program fragments and executable pre- and postconditions. This way, automata get part of the programming language. The resulting language called C++STD contains concepts of design and implementation and can therefore be viewed as a very-high level programming language. For C++STD a prototypical implementation exists.

## 6 CONCLUSIONS

We have presented an automata model for the design phase of object-oriented software engineering methods. A denotational and an operational semantics have been given, which correspond with each other. Because the denotational semantics was based on stream processing functions, the refinement techniques of stream processing functions could be used to define refinement rules for our automata.

The example of a development process has shown that the presented set of refinement rules is simple, flexible and powerful. Therefore, the refinement rules seem to be well-suited for integration in commercial software engineering methods- and tools.

### Thank

We thank Klaus Bergner and Manfred Broy for discussing ideas with us and Bernhard Schätz for a thorough reading of this paper.

# A  STREAMS AND STREAM PROCESSING FUNCTIONS AS A MODEL OF INTERACTIVE SYSTEMS

Stream processing function provide an abstract model for information processing systems and their components. Objects are modeled as components communicating asynchronously with their environment by the exchange of messages. Objects have an *input port* for receiving messages from their environment, and an *output port* for sending messages to their environment.

## A.1  Streams

In our model, the behavior of an object is modeled by its runs, which describe the relationship between the messages arriving at the input port of the object and the messages sent on the output port of the object. We assume that for each run the events on a port are totally ordered, which means that for two different events always one temporarily precedes the other. This allows to model the communication history on a port by a stream of messages.

A *stream* is a finite or infinite sequence of messages. If $M$ denotes the set of messages, $M^*$ the set of all finite sequences of messages and $M^\infty$ the set of all infinite sequences of messages, for the set of all streams over $M$, denoted by $M^\omega$, the equation

$$M^\omega = M^\infty \cup M^*$$

holds.
We will use the following operations and relations:

- $\hat{\ } : M^\omega \times M^\omega \to M^\omega$ denotes the concatenation of two streams, i.e. the stream which is obtained by putting the second argument after the first. The operator $\hat{\ }$ is usually written in infix notation. We assume that

  $$s \in M^\infty \Rightarrow s\hat{\ }t = s,$$

  holds, i.e. the concatenation of an infinite stream $s$ with a stream $t$ yields the stream $s$. $\hat{\ }$ will also be used to concatenate a single message with a stream.

- $\# : M^\omega \to (Nat \cup \{\infty\})$ delivers the length of the stream as a natural number or $\infty$, if the stream is infinite.

- $Filter : \mathcal{P}(M) \times M^\omega \to M^\omega$ denotes the filter-function. $Filter(N, s)$ deletes all elements in $s$ which are not contained in set $N$.

- $ft : M^\omega \to M$ delivers the first element of a stream if the stream has at least one element, and is undefined if its argument is the empty stream.

- $rt : M^\omega \to M^\omega$ removes the first element of a stream if the stream has at least one element, and is undefined if its argument is the empty stream.

- $\sqsubseteq : M^\omega \times M^\omega \to Bool$ is the prefix order between streams. $m \sqsubseteq n$ holds if there exists an $u$ such that $m\hat{\ }u = n$.

Using streams, the communication history of an object can be represented by a pair of streams of messages, where the first component represents the input history of the object and the second component represents the output history of the object.

## A.2  Stream processing functions

The *behavior* of an object is modeled by a *stream processing function* mapping a stream of input messages to a stream of output messages:

$$Behavior : M^\omega \to M^\omega$$

However, not every function with this functionality represents an adequate model of an object: In reality, it is impossible that at any point of time the output depends on future input. To model this fact, we impose an additional mathematical requirement. We require stream processing functions to be *monotone* with respect to to the prefix ordering on streams:

$$x \sqsubseteq y \Rightarrow Behavior(x) \sqsubseteq Behavior(y)$$

An additional requirement, *continuity*, has also to be imposed on stream processing functions as models of objects. We will not define continuity here, but refer to [BDDFGW93]. All monotone and continues stream processing functions are denoted by the function arrow $\xrightarrow{s}$.

While *one* stream processing function can be used to model a *deterministic* agent, we also have to take into account *nondeterminism*. Nondeterminism occurs during the development process due to *underspecification*, or during the lifetime of an agent due to *non-deterministic choice* during the execution. In our model, non-deterministic agents are modeled by *sets of stream processing functions*.

Please note that all the above definitions can easily be extended to objects with more than one input- or output port, see [RKB95] or [BDDFGW93]. While the model of stream processing functions could only be sketched here, in these papers also more details are given.

**REFERENCES** 287